\documentstyle[aps,prl,multicol]{revtex}
\begin{document}


\title{Quantum Dissipation versus Classical Dissipation
for Generalized Brownian Motion}
\author{Doron Cohen}

\date
{
October 1996
\footnote{Published in Phys. Rev. Lett. {\bf 78}, 2878 (1997).}  
}

\address
{
Department of Physics of Complex Systems, \\ 
The Weizmann Institute of Science, Rehovot 76100, Israel. \dag
} 
\maketitle


\begin{abstract}
We try to clarify what are the genuine quantal effects that 
are associated with generalized Brownian Motion (BM). 
All the quantal effects that are associated with the 
Zwanzig-Feynman-Vernon-Caldeira-Leggett model are (formally) 
a solution of the classical Langevin equation. Non-stochastic, 
genuine quantum mechanical effects, are found for a model 
that takes into account either the disordered or the chaotic 
nature of some environment.       
\end{abstract}

\begin{multicols}{2}

The motion of a particle in 1-D, under the influence of an 
environment is commonly described in the classical literature 
by an appropriate generalization of Langevin equation 
\cite{general,gross,CL}
\begin{eqnarray} \label{e1}
m\ddot{x}+\eta\dot{x}={\cal F} 
\end{eqnarray}
Here $m$ and $\eta$ are the 
mass of the particle and the friction coefficient respectively. 
Implicit is an ensemble average over realizations of the 
random force ${\cal F}$. In the standard Langevin equation it 
represents stationary ``noise'' which is zero upon averaging,
and whose autocorrelation function is
\begin{eqnarray} \label{e2}
\langle{\cal F}(t){\cal F}(t')\rangle=\phi(t{-}t') \ .
\end{eqnarray}
This phenomenological description 
can be derived formally from an appropriate Hamiltonian 
${\cal H}={\cal H}_0(x,p)+{\cal H}_{env}$, where the latter  
term incorporates the interaction with environmental degrees 
of freedom. The reduced dynamics of the system may be described 
by the propagator ${\cal K}(R,P|R_0,P_0)$ of the probability 
density matrix. For sake of comparison with the classical limit 
one uses Wigner function $\rho(R,P)$ in order to represent the   
latter. In some cases, using Feynman-Vernon (FV) formalism \cite{FV}, 
an exact path-integral expression for the propagator is 
available \cite{CL}. The FV expression is a double sum 
$\int\int{\cal D}x'{\cal D}x''$ over the path variables 
$x'(\tau)$ and $x''(\tau)$. It is convenient to use 
new path variables $R=(x'{+}x'')/2$ and  $r=(x''{-}x')$, 
and to transform the $\int\int{\cal D}R{\cal D}r$ integral   
into the form \cite{dld}
\begin{eqnarray} \label{e3}  
{\cal K}(R,P|R_0,P_0) \ \ = \ \ 
\int_{R_0,P_0}^{R,P} {\cal D}R \ \ {\cal K}[R]  \ \ \ \ ,
\end{eqnarray}
where ${\cal K}[R]$ is a real functional, which is defined 
by the expression: 
\begin{eqnarray}  \label{e4}  
{\cal K}[R] =
\int {\cal D}r 
\ \ \mbox{e}^{i\frac{1}{\hbar}(S_{free}+S_F)} 
\ \ \mbox{e}^{-\frac{1}{\hbar^2}S_N}
\ \ \ \ \ .
\end{eqnarray}
The ${\cal D}r$ integration is unrestricted at the endpoints,
and the free action functional is 
$S_{free}[R,r]=-m\int_0^td\tau\ \ddot{R}r$. 
The action $S_F[R,r]$ corresponds to the friction  
and the action $S_N[R,r]$ corresponds to the noise. 
The latter are in general non-local  
functionals of the path-variables (there may be long-time  
interactions between different paths segments).  
Still, in practice, it is desired to find a master equation 
of the form 
\begin{eqnarray} \label{e5}
\frac{\partial\rho}{\partial t}={\cal L}\rho 
\end{eqnarray}
that generates essentially the same dynamical behavior. 
Alternatively, it is desired to find an appropriate Langevin 
equation of the form (\ref{e1}), that reproduces the 
reduced dynamics in phase-space.

At this stage it is appropriate to gather few questions
that are of conceptual significance:
{\ \em (a)} What are the essential ingredients that define generic 
generalized Brownian motion (GBM);   
{\ \em (b)} What are the necessary requirements on 
${\cal H}_{env}$ for having generic GBM; 
{\ \em (c)} If ${\cal H}_{bath}$ is strongly chaotic, what is the
minimal number of degrees of freedom which 
are required;
{\ \em (d)} Is it possible to reproduce any generic GBM by assuming
a coupling to an appropriate bath that consists of 
(infinitely many) oscillators;
{\ \em (e)} Is it possible to reproduce any generic GBM by an 
appropriate master equation;
{\ \em (f)} Is it possible to reproduce any generic GBM by an 
appropriate Langevin equation;
{\ \em (g)} In the latter case, what is the relation between 
the noise and the friction, should fluctuation dissipation theorem
be modified ? 
Most frequently questions b and c are emphasized. This  
letter intends to introduce partial answers to the rest, 
less emphasized questions. 

{\em Classically}, the answers for all the questions a-g is known
\cite{yar,wilk}. Any generic ${\cal H}_{env}$ leads to a 
simple BM that is described by (\ref{e1}) with friction which 
is proportional to velocity and white noise 
$\phi(\tau){=}2\eta k_BT \delta(\tau)$ in consistency with 
the classical fluctuation-dissipation theorem.
The environment should consist of at least $3$ degrees 
of freedom with {\em fast} chaotic dynamics.  {\em Fast} 
implies that the classical motion is characterized by 
a continuous spectrum with high frequency cutoff, such 
that the motion of the environment can be treated 
adiabatically with respect to the {\em slow} motion of 
the system. One can use a bath that consist of infinitely
many oscillators in order to reproduce (\ref{e1}). Note that 
an oscillators-bath is obviously {\em non-generic} since it 
consists of non-chaotic degrees of freedom.  Thus, the 
spectral distribution of the oscillators should be 
chosen in a {\em unique} way that mimics the generic 
spectral function and consequently reproduces the simple 
BM behavior.

We are interested in this letter in {\em quantized} BM. 
Just in order to be consistent with the terminology 
that prevails at the literature, we shall use the 
notion ``BM model'' in a restricted sense as referring 
to the quantization of (\ref{e1}) with (\ref{e2}). 
The notion ``GBM'' suggests that a satisfactory model 
should generate {\em additional} physical effects. 
Referring to question a, let us try to list the 
ingredients that should be associated with GBM: 
{\ \em (I)} Fluctuations due to ``noise''; 
{\ \em (II)} Dissipation of energy due to friction effect;
{\ \em (III)} Dissipative diffusion due to competition between 
friction and noise; 
{\ \em (IV)} Non-dissipative diffusion due to ``random walk''
dynamics;
{\ \em (V)} Quantum localization due to quenched disorder;
{\ \em (VI)} Destruction of coherence due to dephasing. 
This list intends to make distinction between  
qualitatively different effects that are   
associated with the {\em reduced} dynamics, irrespective  
of the actual mechanism which is responsible to them.
 
Quantum mechanically it would be desired to derive first,
as in the classical theory, a general description of BM, 
and only later to address question d. However, this turn 
to be impossible, unless uncontrollable approximations 
are made. Therefore we shall take the other way around.   
Referring to question d, it is natural to discuss first 
the the standard model for BM, where  
linear coupling to a large set of harmonic oscillators 
is assumed \cite{gross}. This model has been used extensively 
in the literature. Caldeira and Leggett (CL) and followers \cite{CL}  
have used it to analyze ``Quantum BM'' that corresponds 
to (\ref{e1}) with (\ref{e2}). There,  
in the limit of high temperatures, $\phi(\tau){=}2\eta k_BT \delta(\tau)$
which coincides with the classical limit. 
The friction action functional is 
\begin{eqnarray}   \label{e6}
S_F=-\eta\int_0^t d\tau \ \dot{R}r \ \ ,
\end{eqnarray}
and the noise functional is 
\begin{eqnarray}   \label{e7}
S_N[R,r]=\frac{1}{2} \int_0^t\int_0^t d\tau_1 d\tau_2 
\ \phi(\tau_2{-}\tau_1) \ r(\tau)r(\tau')
\end{eqnarray}
In the absence of noise the ${\cal D}r$ integration is
easily performed leading to 
${\cal K}[R]=\prod_{\tau}\delta(m\ddot{R}+\eta\dot{R})$.
Furthermore, FV have observed \cite{FV} that $S_N$ can be
interpreted as arising from averaging over the realizations
of the classical c-number random force ${\cal F}$.  
Thus, the reduced dynamics of the particle can be reproduced by 
the classical Langevin equation (\ref{e1}) with appropriate $\phi(\tau)$.
Note however that $\phi(\tau)$ will depend on $\hbar$ in 
accordance with fluctuation-dissipation theorem.
In particular, the suppression of ``normal'' diffusion at 
low temperatures \cite{CL} can be interpreted as arising 
from negative noise-autocorrelations \cite{CF}. Also the relaxation 
of a quantal harmonic oscillator to its ground state, 
the dynamics of quantal parametric oscillator with
dissipation, and the dynamics of the quantal kicked rotator 
with dissipation can be simulated by assuming the same 
type of noise (the latter case is analyzed in \cite{ohmic}).   
There is a somewhat more transparent way to observe that the FV-CL 
path integral expression is formally identical with its classical limit 
(for given $\phi(\tau)$).   
With (\ref{e6}) and (\ref{e7}) expression (\ref{e4}) 
for ${\cal K}[R]$ becomes invariant under the replacement 
$\hbar\rightarrow\lambda\hbar$. This replacement is
compensated by the scaling transformation 
$r\rightarrow\lambda r$ of the auxiliary path-variable.

The observation, that ``quantum BM'' 
(in the restricted sense discussed above) 
is formally 
equivalent to the solution of a classical Langevin equation
with colored noise, is probably not new, though there is no 
obvious reference for it. This is probably the reason for 
the existence of extensive literature which utilize rather 
lengthy ``quantum mechanical formalism'' in order to derive 
essentially classical results. 
However, there is a deeper reason for considering ``quantum dissipation'' 
as distinct from ``classical dissipation'' which is concerned with the 
extensive usage of the master equation approach. In this approach the 
commonly used Markovian approximation generates ``non-classical'' 
correction. It is frequently left either unnoticed or unclarified, 
as in a recent publication \cite{recent}, that the resultant non-classical 
feature is an {\em artifact of the formalism} rather than of the model itself.    
In the appendix this point is illustrated by considering a specific
example.

The standard BM motion that is modeled by (\ref{e1}) with (\ref{e2}) 
is not rich enough in order to generate
effects that are associated with the possibly disordered nature
of the environment (ingredients IV and V).  In \cite{dld} we 
have introduced a unified model for the study of diffusion 
localization and dissipation (DLD). The DLD model is defined in 
terms of the path-integral expression (\ref{e3}) with 
\begin{eqnarray}   \label{e9}
S_F \ = \ \eta \int_0^t d\tau \ w'(r(\tau)) \ \dot{R}(\tau) 
\end{eqnarray}
for ohmic friction. The general expression for the noise action 
functional is
\begin{eqnarray}  
S_N[x',x'']=\frac{1}{2} \int_0^t\int_0^t d\tau_1 d\tau_2 
\ \phi(\tau_2{-}\tau_1) \ \ \  \nonumber \\
\label{e10} 
\ \ \ [w(x_2''{-}x_1'')+w(x_2'{-}x_1')-2w(x_2''{-}x_1')]
\end{eqnarray}
where $x_i$ is a short notation for $x(\tau_i)$.
(for white noise see the simplified expression 
(\ref{e10_1}) later).
Both functionals depend on the normalized spatial-autocorrelation 
function $w(x{-}x')$ of the disordered environment. For definiteness 
we have assumed
\begin{eqnarray}  \label{e11} 
w(x{-}x')=\ell^2
\exp\left(-\frac{1}{2}\left(\frac{x{-}x'}{\ell}\right)^2\right) \ .
\end{eqnarray}
The various derivations of the DLD model are discussed 
in \cite{dld}.  Here we note its various limits:
{\ \em (A)} In the classical limit it constitutes a formal solution
of (\ref{e1}) where ${\cal F}(x,t)=-{\cal U}'(x,t)$ and
\begin{eqnarray} \label{e12} 
\langle{\cal U}(x,t){\cal U}(x',t')\rangle=\phi(t{-}t')\cdot w(x{-}x') \ ;
\end{eqnarray}
{\ \em (B)} In the limit $\ell\rightarrow\infty$ it reduces to the 
standard BM model; 
{\ \em (C)} By dropping the friction functional $S_F$ one obtains 
the case of non-dissipative noisy disordered environment; 
{\ \em (D)} By further taking the limit of $\phi(\tau){=}const$ one 
obtains the case of quenched disorder.       
    
The {\em classical} DLD model is similar to the BM model 
for {\em short-time correlated noise}, such that
$\phi(\tau)$ can be approximated by a delta function 
(white noise approximation).  However, for short-time 
correlated noise with negative correlations 
($\int_0^{\infty}\! d\tau \phi(\tau) {\rightarrow} 0$) 
one cannot avoid considering the interplay with the disorder.
This is the case of ``superohmic'' noise and also of low-temperature
``ohmic noise''. In the latter case 
$\phi(\tau){=}-({C}/{\pi})({1}/{\tau^2})$ for $\tau_c{<}\tau$ where
$\tau_c$ is a very short time scale, and $C{=}\hbar\eta$. 
For BM (no disorder) the spatial spreading is 
$\sigma_{spatial}{\sim}(({C}/{\eta^2})({2}/{\pi})\ln t)^{1/2}$
with Gaussian profile, while for the DLD model in the 
same circumstances \cite{dld}
\begin{eqnarray}    \label{e13}
{\cal K}(R|R_0) \ = \ \ 
const \cdot \exp\left( 
-\frac{|R-R_0|}{\left[
4\sqrt{\frac{2}{\pi}} \ \frac{1}{\eta^2 \ \ell} 
\ C \right]}   \right)
\end{eqnarray}
Here $P_0=0$ and an integration over the final $P$ has been performed.
Note that there is a smooth crossover from the BM logarithmic ``diffusion''
(faint noise, dispersion on scale less than $\ell$) to the DLD frozen
profile (stronger noise, dispersion on scale larger than $\ell$).
The classical DLD model becomes significantly distinct from the BM model 
for {\em long-time correlated noise}. In particular, in the limit of 
quenched disorder, the motion of the particle is bounded. More 
generally, for long but finite time autocorrelations, or for higher
dimensionality, the particle will execute non-dissipative 
``random walk'' diffusion.

The {\em quantal} DLD model, in contrast with the ``quantal'' BM model, 
does not constitute a formal solution of its corresponding Langevin 
equation. This leads to some {\em new} genuine quantal effects.
Referring first to the limiting case D of {\em quenched disorder}, 
one may demonstrate that localization is a natural consequence of the 
path-integral expression \cite{dld}. One should wonder whether such 
an effect can be generated by a classical Langevin equation
with appropriate colored noise.  The frozen ``diffusion'' profile
(\ref{e13}) is probably the best that one can achieve.  
However, the reduced dynamics is not the same as in the case of 
quantum localization, since there is a strong velocity-position 
correlation. Thus, it is claimed that quantal localization cannot be 
generated by a classical Langevin equation.

There is an additional aspect of the quantal DLD model that cannot 
be generated by a classical Langevin equation. {\em The distinction 
between the quantal DLD model and its classical limit persists  
even in the limit of high temperatures.}  In order to 
clarify this point one should substitute $\phi(\tau){=}2k_BT\delta(\tau)$
into (\ref{e10}) yielding
\begin{eqnarray}    \label{e10_1}
S_N[r] = 2\eta K_BT \int_0^t \ [w(0){-}w(r(\tau))] \ d\tau \ ,
\end{eqnarray}
and compare (\ref{e9}) and (\ref{e10_1}) with their classical limit, 
which is not by accident (\ref{e6}) and (\ref{e7}) respectively.  
The quantal expressions \cite{new} differ from the classical ones 
for $\ell\ll |r|$. The scaling properties of $r$ with $\hbar$ imply
that these large deviations are important for the study 
of interference and dephasing. Simply by inspection of the 
action functionals, one may draw two important observations:
First, interference in the DLD model is not affected by friction, 
unlike BM model. The second observation is that the dephasing 
factor is
\begin{eqnarray} \label{e14}  
\langle \mbox{e}^{i\varphi} \rangle \  = \ 
\mbox{e}^{-S_N[\ell\ll|r|]} \ = \
\exp\left[-\frac{2\eta k_BT\ell^2}{\hbar^2} \cdot t \right]  \ , 
\end{eqnarray}
irrespective of the geometry of the interfering paths $x_a(\tau)$
and $x_b(\tau)$ which are assumed to be well separated with respect
to the microscopic scale $\ell$ (above $r=(x_a{-}x_b)$). 
The latter conclusion should be contrasted 
with the BM case where 
\begin{eqnarray}  \label{e15}  
\langle \mbox{e}^{i\varphi} \rangle \  = \ 
\exp\left[-\frac{1}{2}\frac{2\eta k_BT}{\hbar^2}\int_0^t 
\ (x_a(\tau){-}x_b(\tau))^2 \ d\tau \right] \ . 
\end{eqnarray}
which is essentially the same as the dephasing due to the  
interaction with extended (electromagnetic) field modes \cite{stern}. 
Thus, dephasing due to the interaction with disordered environment 
(e.g. localized impurities) is qualitatively different.
Further discussion, semiclassical considerations and 
specific examples will appear in \cite{dld}. In particular,  
it is interesting to note that due to interference, the familiar 
diffusive behavior is modified by a ballistic component that decays 
exponentially in time as in (\ref{e14}).     

Finally, we should refer to question d which is also 
intimately related to question g concerning the role
of fluctuation-dissipation theorem. One should ask, 
whether the DLD model is 
the ``ultimate'' model for the description of BM
in the most generalized way (as far as generic 
effects are concerned). 
In case of 2-D generalized BM one should consider 
also the effect of ``geometric magnetism'' \cite{berry},
which is not covered by the 1-D DLD model.
Here we limited the discussion to 1-D BM.  
In order to answer this 
question one should consider a general 
nonlinear coupling to a thermal, possibly 
chaotic bath. In the limit of weak coupling  
one may demonstrate \cite{dld} that indeed
the bath can be replaced by an equivalent 
``effective bath'' that consists of harmonic 
oscillators, yielding the DLD model.
In the opposite limit of 
strong coupling, and extremely adiabatic interaction, the 
reduced dynamics is determined by the ground state
energy $E_{env}(x)$ of ${\cal H}_{env}$, leading to 
an effective ``quenched'' disordered potential.
Such extreme adiabaticity is probably not very realistic. 
Gefen and Thouless \cite{efrat}, Wilkinson \cite{wilk} 
and Shimshoni and Gefen \cite{efrat} have emphasized 
the significance of Landau-Zener transitions as a
mechanism for dissipation. 
There is a possibility that some future 
derivation, will demonstrate 
that an equivalent ``oscillators bath'' can 
be defined also in this case.
The existence of such derivation is most 
significant, since it implies that no ``new effects''
(such as ``geometric magnetism'' in case of 2-D 
generalized BM) can be found in the context of 
1-D generalized BM.
Wilkinson has demonstrated that due to the Landau-Zener mechanism anomalous
friction, which is not proportional to velocity,
arise for GOE fermion bath \cite{wilk}. 
The BM model cannot generate such anomalous effect, due to 
a ``memory problem'' that makes it ill-defined. However, 
one may demonstrate that the {\em non-ohmic} DLD model
can be used in order to generate this effect \cite{dld}.

{\em Appendix} - Here we shall illustrate how an apparently 
non-classical feature may arise due to the application of the 
Markovian approximation.
To demonstrate this point in a transparent way it is best to make a
reference to a related recent study \cite{hanggi,kohler} of 
the the parametric driven harmonic quantum oscillator with ohmic 
dissipation. This problem has an exact solution using FV formalism 
\cite{CF,hanggi}.  In \cite{kohler} various approximation schemes 
for ${\cal L}$ in (\ref{e5}) has been discussed, leading to an expression
of the general form 
\begin{eqnarray} \label{e8}
{\cal L}=-\frac{p}{m}\partial_x+\frac{\eta}{m}\partial_pp+...
+D_{pp}\partial_p^2+D_{xp}\partial_x\partial_p \ .
\end{eqnarray}
(The time-dependent driving term has been omitted for brevity).
The last term is the so-called ``Drude correction''. Due to this
term the diffusion matrix is no longer positive semidefinite. 
Kohler et al. \cite{kohler} have correctly pointed out that 
consequently ${\cal L}$ has no equivalent Langevin representation. 
Due to this term  Wigner function may become negative in some 
places in phase space. Note however that (\ref{e8}) is the best 
approximation for the actual dynamics within the framework of 
the Master equation approach.
We shall demonstrate now that the ``Drude correction''
may be derived in a very simple way from the classical Fokker-Planck 
equation. This derivation also sheds new light on the traditional Markovian  
approximation which is used within the framework of the master equation 
approach.  Starting from (\ref{e1}) with a definite realization of the
random force $F$, Liouville equation is 
$\frac{\partial\rho}{\partial t}=-\nabla(\rho v)$ where 
$\nabla=(\partial_x,\partial_p)$ and $v=(p/m,F_{total})$
with $F_{total}=-\eta{\cdot}p/m - {\cal F}(t)$.  The first two terms in
(\ref{e8}) are immediately obtained, and the additional term due to 
the random force is $-{\cal F}(t){\cdot}\partial_p\rho$. 
We now use the identity $\rho|_{\cal F}(x(t)|_{\cal F},p(t)|_{\cal F},t)=
\rho|_{{\cal F}=0}(x(t)|_{{\cal F}=0},p(t)|_{{\cal F}=0},t)$, which holds
since both sides equals $\rho(x(0),p(0),t{=}0)$. One substitutes 
$x(t)|_{\cal F}=x(t)|_{{\cal F}=0}+\int_0^t G(t,\tau){\cal F}(\tau)d\tau$,
where $G$ is the appropriate Green function (response kernel) of (\ref{e1})
with parametric driving term that should be included. Expanding 
$\rho$ with respect to ${\cal F}$ up to first order, and then averaging 
$-{\cal F}(t){\cdot}\partial_p\rho$ over realizations of the random force, 
one obtains the last two terms in (\ref{e8}). In particular, 
the Drude term is $D_{xp}=\int_0^t\phi(t{-}\tau)G(t,\tau)d\tau$. 
It is easy to observe that this result coincides with Eq.(85) of
\cite{kohler}. Evidently, in the 
high temperature limit (white noise) this term goes to zero.
However, at the limit of zero temperature $\phi(\tau)$ constitutes
a Fourier transform of $\phi(\omega)=\eta\hbar|\omega|$ in accordance
with fluctuation-dissipation theorem, leading to diffusion matrix 
that is no longer positive semidefinite.
Thus, we have demonstrated that the ``Drude correction'' does not imply 
that the exact quantum dynamics cannot be generated by an 
appropriate Langevin equation, rather it is an artifact of the 
Markovian approximation involved.

I thank Shmuel Fishman from the Technion for supporting 
and encouraging the present study in its initial phases.  
I also thank Harel Primack and Uzy Smilansky for interesting 
discussions and suggestions, and Yuval Gefen for his comments. 
The research reported here was supported in part by  the 
Minerva Center for Nonlinear Physics of Complex systems.


\end{multicols}
\end{document}